\documentclass[aps,prd,reprint,showpacs,superscriptaddress,groupedaddress]{revtex4-1}

\usepackage{times}
\usepackage{graphicx,bm,amssymb}
\usepackage{ulem}
\usepackage{epsfig}
\usepackage{natbib}
\usepackage{psfrag}
\usepackage{braket}
\usepackage{hyperref}
\usepackage{dcolumn}
\usepackage{color}

\def\araa{ARA\&A}
\def\apj{ApJ}
\def\apjl{ApJ}
\def\apjs{ApJS}
\def\mnras{MNRAS}
\def\prd{Phys.~Rev.~D}
\def\pasp{PASP}
\def\nat{Nature}
\def\physrep{Phys.~Rep.}

\newcommand{\msun}{\mbox{$M_\odot$}}

\def\be{\begin{eqnarray}}
\def\ee{\end{eqnarray}}
\def\bi{\begin{itemize}}
\def\ei{\end{itemize}}
\def\lsim{\mathrel{\rlap{\lower3pt\hbox{\hskip1pt$\sim$}}
     \raise1pt\hbox{$<$}}} 
\def\gsim{\mathrel{\rlap{\lower3pt\hbox{\hskip1pt$\sim$}}
     \raise1pt\hbox{$>$}}} 

\newcommand{\bary}{\begin{eqnarray}}
\newcommand{\eary}{\end{eqnarray}}

\begin{document}

\title{Neutrino Signal from Compact Objects during  their Formation, their Mergers,\\ or as a Signature of Electric-Charge Phase Transition}
\author{Nissim Fraija}  
\email[]{nifraija@astro.unam.mx}
\affiliation{Instituto de Astronom\'ia, Universidad Nacional Auton\'oma de M\'exico,  Mexico D.F. 04510, Mexico,}
\author{Enrique Moreno M\'endez}
\email[]{enriquemm@ciencias.unam.mx}
\affiliation{Facultad de Ciencias, Universidad Nacional Auton\'oma de M\'exico,  Mexico D.F. 04510, Mexico,}
\author{Gibr\'an Morales}  
\email[]{gmorales@astro.unam.mx}
\affiliation{Instituto de Astronom\'ia, Universidad Nacional Auton\'oma de M\'exico,  Mexico D.F. 04510, Mexico.}
\author{Alfredo Saracho}  
\email[]{asaracho@astro.unam.mx}
\affiliation{Instituto de Astronom\'ia, Universidad Nacional Auton\'oma de M\'exico,  Mexico D.F. 04510, Mexico.} 
\date{\today}





\begin{abstract}

We study neutrino production, propagation, and oscillations within an extremely magnetized background of finite-temperature nuclear matter.  We focus on three particularly interesting cases and identify the astrophysical scenarios where such a signal may be found.   The first case involves nuclear matter with electrons, and it is found during the central-engine stage of, both, short and long gamma-ray bursts (GRBs). Thus, for the short GRB case it will also be associated to gravitational-wave events where there is an electromagnetic counterpart (e.g., GW170817). 
The second and third scenarios involve the presence of strange-quark matter (SQM).  The second scenario occurs if SQM can become negatively charged (SQM$^-$; which may only occur at high pressure) and, thus, it is embedded in a positron plasma.
The third case may be found at the interphase where SQM transitions from positive (SQM$^+$) to negative; here, positrons and electrons may constantly annihilate and give a distinctive neutrino signature.  Therefore, this may also be a signature of the existence of strange stars.
Given the wide range of magnetic fields we find in the literature, we also briefly discuss the maximum limit that a stellar mass compact object may posses.
\end{abstract}

\maketitle\thispagestyle{empty}



\section{Introduction}

Massive stars (those with Zero-Age-Main-Sequence mass $M_{\rm zams} \gtrsim 8 \msun$) evolve until a large iron core is produced.   
When the iron core is depleted of nuclear fuel (mostly Si and S) it can no longer sustain hydrostatic equilibrium and collapses. 
The result, if the Fe core is not too massive, is a core-collapse supernova (CC-SN), likely driven by neutrinos \citep{1979NuPhA.324..487B,1990RvMP...62..801B}, as well as the formation of a compact object (CO; e.g. neutron stars, strange stars, or black holes; which we will refer to as NSs, SSs, or BHs).

Unlike the simplest case of CC-SNe, Gamma-ray bursts (GRBs) require a central engine where a compact object rapidly rotates, and an accretion disk forms in its vicinity.  It is also necessary to have extremely strong magnetic fields (which are a likely result of differential rotation and convection and/or conservation of magnetic flux) mediating the interaction between compact object and surrounding material (e.g., Blandford-Znajeck mechanism \citep{1977MNRAS.179..433B,2007ApJ...671L..41B} or magnetar model \citep{1994MNRAS.270..480T, 2011MNRAS.413.2031M}).  Temperatures must be of the order of $10^9$ to $10^{11}$ K, and, thus, large numbers of neutrinos and antineutrinos are produced and, given symmetry considerations, they may annihilate (preferentially) along the rotational axis and help produce extremely energetic, relativistic jets which we observe as GRBs.
Now, these conditions ($T$, B field, $\rho$, and $P$) are likely present during CC-SNe for rapidly rotating Fe cores (collapsar \citep{1993ApJ...405..273W, 1999ApJ...524..262M} and/or magnetar models \citep{1994MNRAS.270..480T, 2011MNRAS.413.2031M}) as well as for compact object mergers \citep{2014ARA&A..52...43B, 2016ApJ...831...22F}. Thus we will work with both of these scenarios.

The equation of state (EoS) of nuclear matter is not well known as Quantum Chromodynamics (QCD) is not helpful in estimating the properties of large-density, low-temperature nuclear matter \citep[see, e.g., discussion in][]{2000csnp.conf.....G}.  
Instead, phenomenological (or semiempirical) theories have been proposed in order to describe the possible properties of nuclear matter at large density \citep[see][]{1974PhRvD...9.3471C,1961PhRv..122..345N,1961PhRv..124..246N}.

Shortly after the quark model was proposed, it was suggested that stars with free quarks could form \citep{1970PThPh..44..291I}. 
\citet{Witten:1984rs}  suggests that such stars would likely form if Strange-Quark Matter (SQM) were the ground state of nuclear matter at large nucleon number \citep[previously hinted for heavy-ion collisions in][]{Terazawa:1979hq}.
In \citet{1984PhRvD..30.2379F,1986ApJ...310..261A}, and \citet{Haensel:1986qb}, SQM is studied using the MIT bag model \citep[for a review on this topic see][]{2000csnp.conf.....G}.
They used the first order (in the strong coupling constant, $\alpha_s$) calculation by \citet{1979PhRvD..20.2353B} which usually gives low-maximum-mass ($M_{SS,MAX}\lesssim 2 \msun$) for Strange Stars (SSs).
However, in the last few years, compact stars have been found with masses of $1.97 \msun$ \citep{2010Natur.467.1081D} and $2.01 \msun$. 
Nonetheless, \citet{2010PhRvD..81j5021K} recently performed second order (O$(\alpha_c^2)$) estimates of the maximum mass for SSs finding numbers on the order of $2.75 \msun$.  Thus, the existence of SSs has not been excluded \citep[see, e.g.,][for further discussion on this topic]{2007Natur.445E...7A}. 

SQM was first studied at finite temperature by \citet{1988PhLB..202..133R}, \citet{1989PhRvD..39.1233A}, and \citet{1989PhRvD..40..165C}.  They found that SQM may be stable for $T \lesssim (20$ -- $30)$ MeV (where 1eV$\simeq 1.1\times10^{4}$ K); thus, we expect that a few seconds after core collapse the proto-neutron star (NS) could transition into a hot SS.  In fact, \citet{Fischer2018} have recently performed a study on CC-SNe with a QM (without strange quarks) phase transition that rejuvenates the SN shock a few seconds ($\sim 3$ s) after the initial rebound.

Another scenario where a large-temperature SQM plasma may be found is during the merger of a SS with another compact object (SS-SS, NS-SS, or BH-SS).  A collision of two compact objects (other than BHs) has recently been observed during the high-energy transient event GW/GRB 170817 \citep{PhysRevLett.119.161101, 2041-8205-848-2-L12}.  

Strong magnetic fields, as large as $10^{16}$ G have been estimated on magnetars \citep{2017ARA&A..55..261K}. Indeed, the Collapsar model \citep{1993ApJ...405..273W} with the aid of, either, the Blanford-Znajek mechanism \citep{1977MNRAS.179..433B,2002ApJ...575..996L,2011ApJ...727...29M}, or the Magnetar model \citep{2011MNRAS.413.2031M}, requires extremely large magnetic fields ($B\gtrsim10^{14}$ G)  to produce long GRBs \citep{2016ApJ...818..190F,2015ApJ...804..105F,2017ApJ...848...15F,2017ApJ...848...94F, 2019arXiv190406976F} and, in particular, those that last thousands of seconds.  
Now, simple estimates using differential rotation and magnetic flux conservation (in perfect magnetohydrodynamics; MHD) during core collapse can show that the internal field of the compact object can be a couple orders of magnitude larger \citep[see, e.g.,][]{2014ApJ...781....3M};  this can also be achieved through dynamos. In principle, it is energetically possible to build internal fields as large as $10^{18}$ G, thus we shall limit this study to that maximum.

Using the MIT-bag model the values for the bag constant ($B_{M}$), the strong-coupling constant ($\alpha_s$) and the mass of the strange quark ($m_s$) can be varied \citep[we make use of the model described in][which still uses O$(\alpha_s)$]{2013arXiv1306.1828M}.  
For values of $\alpha_s\gtrsim0.5$, strange quark matter (SQM) can become negatively charged as the density increases.
And for $\alpha_s \sim 0.9$ negatively charged SQM (SQM$^-$) is stable at zero pressure. 
We know SQM$^-$ at zero pressure could not be stable in nature as compact stars merge and part of their matter is released back into the universe;
if SQM$^-$ were to get in touch with normal matter there would be no Coulomb barrier to prevent normal matter from being converted into SQM$^-$.
We know this is not the case as we do not see stars nor our planet being converted.
Thus, we will assume that if SQM$^-$ exists it is stable only at high pressure and it is surrounded by SQM$^+$ (positively charged SQM).
  
Figure 1 on \citet{2014arXiv1401.3787F} (where $\alpha_s = 0.6$) shows how electrons are replaced by positrons once the baryon density reaches values of $n_{bar} \sim 0.8$ fm$^{-3}$. 
In such case, a region where e$^-$--e$^+$ pairs annihilate into neutrinos may form.   
Charge neutrality has to be locally achieved and, thus, beta equilibrium will have to provide electrons (positrons) where $SQM^+$ ($SQM^-$) exists to replace those annihilated at the $SQM^+$--$SQM^-$ interface.
Hence, extra neutrinos (anti-neutrinos) will be produced by these beta-equilibrium reactions. 
From e$^-$--e$^+$ pair annihilation we know the neutrinos should have $E_\nu \gtrsim 0.511$ MeV, however the equality will only occur for pairs with no Fermi momentum.  
Those at the surface of the Fermi sphere will be the most energetic ones given that the compact star has little thermal energy density. 

Neutrinos provide crucial pieces of information in the three scenarios described above. However, the  properties of these  neutrinos get modified when they propagate in the strongly magnetized medium, and depending on their flavors, they feel a different effective potential.  This occurs because the electron neutrino ($\nu_e$) interacts with electrons via both, neutral and charged currents (CC), whereas muon ($\nu_\mu$) and tau ($\nu_\tau)$ neutrinos interact only via the neutral current (NC).  This would induce a coherent effect in which maximal conversion of $\nu_e$ into $\nu_\mu$ ($\nu_\tau$) takes place even for a small intrinsic mixing angle.  
The resonant conversion of neutrino from one flavor to another due to the medium effect, is well known as the Mikheyev-Smirnov-Wolfenstein effect \citep{wol78}.
In this work, we roughly estimate the number of neutrino events and flavor ratio expected on the current and future neutrino detectors.   For this reason, we calculate the neutrino effective potential and then study the propagation and resonant oscillations of thermal neutrinos in these electron-and-positron, highly-magnetized,  plasmas which may be generated in the three scenarios.  By considering the two-neutrino mixing solar, atmospheric and accelerator parameters  we find that resonant oscillations are strongly dependent on the angle of propagation with respect to the magnetic field ($\varphi$).
We find a strong suppression of neutrino oscillations when the propagation is close to parallel to the magnetic field.  
Finally, we discuss our results in the three described frameworks.

What is relevant in this paper:
\begin{itemize}
\item{The idea of cooling by pair annihilation in a charge-phase transition. Which may also occur in other scenarios, e.g., kaon condensation in neutron stars.}
\item{The angle dependence of the neutrino propagation and magnetic field, which could lead to important field configuration information which is otherwise unavailable to external observers.}
\item{This mechanism should also produce a cooling curve which differs from other curves predicted in the literature (this is an observable through the usual channels, i.e., electromagnetic radiation).  Were this curve to fit the observational data it would reveal important information on the equation of state of nuclear matter at large density.}

\end{itemize}
This paper is arranged as follows: In Section 2 we present the neutrino effective potential for three regions; a transition region, 
SQM$^+$ (only electrons) and SQM$^-$ (only positrons). In section 3 we show the neutrino production and detection. In Section 4, we show the neutrino oscillations. In sections 5 and 6, we present our results, discussion and conclusions.\\
%
%
%
\section{Neutrino  Effective Potential}\label{sec-Justification}

We use the finite-temperature, field-theory formalism to study the effect of a heat bath on the propagation of elementary particles. The effect of magnetic fields is taken into account through Schwinger's propertime method \citep{1951PhRv...82..664S}.   The effective potential of a particle is calculated from the real part of its self-energy diagram. The neutrino field equation of motion in a magnetized medium is
\be
[ {\rlap /k} -\Sigma(k) ] \Psi_L=0\,,
\label{disneu}
\ee
where the neutrino self-energy operator $\Sigma(k)$ is a Lorentz scalar which depends on the characterized parameters of  the medium, as for instance,  chemical potential, particle density, temperature, magnetic field, etc.  Solving this equation and using the Dirac algebra, the dispersion relation $V_{eff}=k_0-|{\bf k}|$ as a function of Lorentz scalars can be written as
\be
V_{eff}=b-c\,\cos\varphi-a_{\perp}|{\bf k}|\sin^2\varphi\,,
\label{poteff1}
\ee
where $\varphi$ is the angle between the neutrino momentum and the magnetic field vector.   Now the Lorentz scalars $a$, $b$ and $c$ which are functions of neutrino energy, momentum and magnetic field can be calculated from the neutrino self-energy 
\be
Re \Sigma_W(k)=R\,[a_{W_\perp} \rlap /k_\perp + b_W \rlap /u + c_W \rlap /b]\,L\,,
\ee
 due to charge current and neutral current interaction of neutrino with the background particles. In a strong magnetic field, the charged particles are confined to the Lowest Landau level ($n=0$ for $\left(\frac{T}{m_e}\right)^2 \ll \left( \frac{B}{B_{\rm c}}\right)^2$), therefore, following \citet{2014ApJ...787..140F}, the Lorentz scalars in natural units ($c=\hbar=k=1$)can be calculated through the total one-loop  neutrino self-energy  in a highly magnetized medium  which is given by  
{\scriptsize
\bary
b_W=&&\sqrt2 G_F\biggl[ \biggl( 1+\frac32\frac{m_e^2}{M_W^2}+ \frac{eB}{M_W^2} +\frac{E_{\nu_e}k_3}{M_W^2}+\frac{E^2_{\nu_e}}{M_W^2}\biggr)(N^0_e - \bar{N}^0_e)\cr
&&\hspace{1cm}  -\frac{eB}{2\pi^2M_W^2}\int^\infty_0\,dp_3\biggl\{2\,k_3E_{e,0}+2E_{\nu_e}\biggl( E_{e,0} - \frac{m_e^2}{2E_{e,0}}\biggr) \biggr\} \cr
&&\hspace{5.6cm} \times (f_{e,0}+\bar{f}_{e,0})       \biggr]\,,
\label{Lescb}
\eary
}
and
{\scriptsize
\bary
c_W=&&\sqrt2 G_F\biggl[ \biggl( 1+\frac12\frac{m_e^2}{M_W^2}+ \frac{eB}{M_W^2} -\frac{E_{\nu_e}k_3}{M_W^2}-\frac{k^2_3}{M_W^2}\biggr)(N^0_e - \bar{N}^0_e)\cr
&&\hspace{2.0cm} -\frac{eB}{2\pi^2M_W^2}\int^\infty_0\,dp_3\biggl\{ 2E_{\nu_e}\biggl( E_{e,0} - \frac{m_e^2}{2E_{e,0}}\biggr)\cr
&&\hspace{2.6cm} +2k_3\biggl( E_{e,0} - \frac{3m_e^2}{2E_{e,0}}\biggr) \biggr\}  (f_{e,0}+\bar{f}_{e,0})       \biggr]\,,
\label{Lescc}
\eary
}
where the number density of electrons can be written as
\be\label{ne}
N_e^0=\frac{eB}{2\pi^2}\int^\infty_0 dp_3 f_{e,0}\,,
\ee
and
\be
f(E_{e,0})=\frac{1}{e^{\beta(E_{e,0}-\mu)} +1}\,,
\ee
with the  electron energy in the lowest Landau level given by
\be
E^2_{e,0}=(p^2_3+m_e^2)\,.
\ee
where $m_e$ is the electron mass, $M_W$ is the W-boson mass, $G_F$ is the Fermi coupling constant, $\beta=T$ and $B_c$ is the critical magnetic field.\\
From figure~3 in \citet{2013arXiv1306.1828M} we know that the electron and positron chemical potentials vary from $\mu_e\simeq15$ MeV at baryon density $n_{\rm bar} = 0.3$ baryon fm$^{-3}$ to $\mu_e\simeq0.5$ MeV at $n_{\rm bar} =10$ baryon fm$^{-3}$. We consider the condition $\mu_e \lesssim $E$_e$,  for which the chemical potential of positrons and electrons is smaller than their energies. In this case,
the fermion distribution function can be written as a sum  given by,
{\small
\be\label{fe}
f(E_{e,0})=\frac{1}{e^{\beta(E_{e,0}-\mu)} +1}\approx\sum^{\infty}_{l=0}(-1)^l e^{-\beta(E_{e,0}-\mu)(l+1)}\, .
\ee
}
It is worth noting that the number density of electrons and positrons computed through Eq. (\ref{ne}) is insoluble for the condition $\mu_e \gtrsim E_e$ (see Appendix A).  Replacing eqs. (\ref{ne}) and (\ref{fe}) into (\ref{poteff1}) and solving the integral-terms in eqs. (\ref{Lescb}) and (\ref{Lescc}), we can calculate the potential for three cases.

\subsection{Transition region, $N^0_e\simeq \bar{N}^0_e$}

The first case occurs at the interface between the region dominated by SQM$^-$ and the region dominated by SQM$^+$, i.e., where the electric phase transition occurs and where e$^-$ -- e$^+$ pairs will annihilate into $\nu_{\rm e}$ -- $\bar{\nu}_{\rm e}$ pairs.
In this region the chemical potentials for electrons and positrons are both zero, $\mu_{\rm e^-} = \mu_{\rm e^+} = 0$.

The effective potential for such a region is given by
{\small
\bary
V_{eff}&=&-\frac{4\sqrt2 G_F\,m_e^4\,E_\nu\,\Omega_B}{\pi^2\,m_W^2}\times \cr
&&\sum^\infty_{l=0} (-1)^l\left[\frac34K_0(\sigma_l)+\frac{K_1(\sigma_l)}{\sigma_l} (1-\cos\varphi) \right]\,,
\eary
}
where $\Omega_B=\frac{B}{B_c}$, $\alpha_l=\beta\mu(l+1)$ and $\sigma_l=\beta m_e(l+1)$. This region is likely a thin spherical shell within the star.

\subsection{SQM$^+$ and electrons, $\bar{N}^0_e\simeq0$}

The outer part of the strange star (should no normal matter lie on its surface supported by the electron layer extending a few hundreds of  Fermi from the quark matter surface) will consist, as discussed above, of SQM$^+$.
Therefore, this mantle will be kept electrically neutral by the presence of electrons.

The effective potential for neutrinos traversing this region will be given by
{\small
\bary
V_{eff}=\frac{\sqrt2 G_F\,m_e^3\,\Omega_B}{2\pi^2}\sum^\infty_{l=0} (-1)^l e^{\alpha_l} (V_a-V_b)
\label{poteff2}
\eary
}
where
{\small
\bary\label{Va}
V_a&=&K_1(\sigma_l)\biggl\{\left(1+ \frac{m_e^2}{m^2_W}\left(\frac32+2\frac{E^2_\nu}{m^2_e} +\Omega_B \right)\right)\cr
&&\hspace{1cm}-\left(1+ \frac{m_e^2}{m^2_W}\left(\frac12-2\frac{E^2_\nu}{m^2_e} +\Omega_B\right)\right)\cos\varphi\biggr\}
\eary
}
and
\be\label{Vb}
V_b= 4\frac{m_e^2}{m^2_W}\frac{E_\nu}{m_e}\left(\frac34K_0(\sigma_l)+\frac{K_1(\sigma_l)}{\sigma_l}(1-\cos\varphi)\right)\,.
\ee
%
%
%
%
\subsection{SQM$^-$ and positrons, $N^0_e\simeq0$}

The core of the strange star in our model with a pressure-induced electric phase transition in the SQM will need positrons to keep local charge neutrality.
In this SQM$^-$ region the effective potential to which crossing neutrinos will be subjected is 
{\small
\bary
V_{eff}=-\frac{\sqrt2 G_F\,m_e^3\,\Omega_B}{2\pi^2}\sum^\infty_{l=0} (-1)^l e^{-\alpha_l} (V_a+V_b)
\label{poteff2}
\eary
}
where $V_a$ and $V_b$ are, again, given by eqs. \ref{Va} and \ref{Vb}.
%
%

\section{Neutrino Flux}
\subsection{Neutrino Production}
Limiting the total mass lost to neutrino cooling of a compact star (NS or SS) to half a solar mass (an overestimate even for the most massive ones) provides us with $E_\nu \sim (10^{33}\, {\rm g}) \sim 10^{54}\,{\rm erg}$.
According to \citet{2004ApJS..155..623P,2011PhRvL.106h1101P} the $\nu-$cooling timescale may be as long as a few 
\be
\tau_\nu \sim 10^6 {\rm yr} \sim \pi\times 10^{13} {\rm s},  
\ee 
thus allowing for a neutrino luminosity of up to 
\be
L_\nu = E_\nu / t_\nu \sim 10^{40} {\rm erg\;\; s}^{-1}.
\ee
This energy and luminosity budgets must be shared between the beta equilibrium reactions between the quarks up ($u$), down ($d$) and strange ($s$) occurring throughout the star:
\begin{itemize}
    \item SQM$^+$:
    \be
    u+e^-\rightarrow d+\nu_e\hspace{1cm}
    d\rightarrow u+e^-+\bar{\nu}_e\nonumber\\
    u+e^-\rightarrow s+\nu_e \hspace{1cm}
    s\rightarrow u+e^-+\bar{\nu}_e\\\nonumber
    \ee
    
    \item SQM$^-$:
    \be
    d+e^+\rightarrow u+\bar{\nu}_e \hspace{1cm}
    u\rightarrow d+e^++\nu_e\nonumber\\
    s+e^+\rightarrow u+\bar{\nu}_e \hspace{1cm}
    u\rightarrow s+e^++\nu_e\\\nonumber
    \ee
\end{itemize}
and $e^\pm$ pair annihilation at the SQM$^+$--SQM$^-$ interface region:
\begin{itemize}
    \item $e^\pm$ pair annihilation:
    \be
    e^++e^-\to \nu_x+\bar{\nu}_x.
    \ee
\end{itemize}   
The subscript x indicates the neutrino flavor: electron, muon, or tau.

We will produce calculations where the neutrino luminosity from the $e^\pm$ pair annihilation at the SQM$^+$--SQM$^-$ interface is a small fraction of the total luminosity ($L_\nu \sim 10^{37}\,$ erg s$^{-1}$) to a large percentage ($L_\nu \sim 10^{40}\,$ erg s$^{-1}$).

\subsection{Neutrino Detection}
It is possible to estimate the number of events expected in current (\textit{Super-Kamiokande}; SK) and future (\textit{Deep Underground Neutrino Experiment}; DUNE and \textit{Hyper-Kamiokande}; HK) neutrino observatories.  Details about the technical specifications of SK, DUNE and HK can be found in \cite{fuk03}, \cite{2014arXiv1412.4673H} and \cite{acc16}, respectively.  Then, the events expected can be written as


\be
N_{ev}=t\,V N_A\,  \rho_N  \int_{E'} \sigma^{\bar{\nu}_ep}_{cc} \frac{dN}{dE}\,dE
\ee

\noindent where $V$ is the effective volume of water, $N_A=6.022\times 10^{23}$ g$^{-1}$ is Avogadro's number, $\rho_N=2/18\, {\rm g\, cm^{-3}}$ is the nucleons density in water, $ \sigma^{\bar{\nu}_ep}_{cc}\simeq 9\times 10^{-44}\,E^2_{\bar{\nu}_e}/MeV^2$  is the neutrino cross section, $t$ is the observed time and $dN/dE$ is the neutrino spectrum. Taking into account the relationship between the  neutrino luminosity $L_{\bar{\nu}_e}$ and flux $F_{\bar{\nu}_e}$, $L_{\bar{\nu}_e}=4\pi d^2_z  F_{\bar{\nu}_e}\braket{E}=4\pi d^2_z   E^2 dN/dE$ and approximation of the time-integrated average energy and time, then the expected event number is given by: 
\bary
N_{ev}&\simeq&\frac{t}{\braket{E_{\bar{\nu}_e}}}V N\,  \rho_N  \sigma^{\bar{\nu}_ep}_{cc} \braket{E_{\bar{\nu}_e}}^2\frac{dN}{dE}\cr
&\simeq&\frac{t}{4\pi d^2_z \braket{E_{\bar{\nu}_e}}}V N\,  \rho_N  \sigma^{\bar{\nu}_ep}_{cc}\,L_{\bar{\nu}_e}\,,
\eary
where $d_z$ is the distance from this hypothetical source.

\section{Neutrino Oscillation}

%
Measurements of fluxes of solar, atmospheric and accelerator neutrinos have shown overwhelming evidence for neutrino oscillations and then for neutrino masses and mixing. To make a full analysis,  we are going to show the important quantities to involve in neutrino oscillations in vacuum and matter as well as the two and three-mixing parameters. The two-mixing parameters are related as follows:\\
\textbf{Solar Experiments}: A two-flavor neutrino oscillation analysis yielded $\delta m^2=(5.6^{+1.9}_{-1.4})\times 10^{-5}\,{\rm eV^2}$ and $\tan^2\theta=0.427^{+0.033}_{-0.029}$\citep{aha11}.\\
\textbf{Atmospheric Experiments}: Under a two-flavor disappearance model with separate mixing parameters between neutrinos and antineutrinos the following parameters for the SK-I + II + III data $\delta m^2=(2.1^{+0.9}_{-0.4})\times 10^{-3}\,{\rm eV^2}$ and $\sin^22\theta=1.0^{+0.00}_{-0.07}$ were found.\citep{abe11a}.\\
\textbf{Accelerator Experiments}: \cite{chu02} found two well-defined regions of oscillation parameters with either $\delta m^2  \approx  7\, {\rm eV^2}$ or $\delta m^2 < 1\, {\rm eV^2} $ compatible with both LAND and KARMEN experiments, for complementary confidence. In addition, MiniBooNE found evidence of oscillations in the 0.1 to 1.0 eV$^2$, which are consistent with LSND results \citep{ath96, ath98}.\\
The solar, atmospheric and accelerator parameters are related in three-mixing parameters as follows \citep{aha11,wen10}:
\bary\label{3parosc}
{\rm for}&&\,\,\sin^2\theta_{13} < 0.053: \delta m_{21}^2= (7.41^{+0.21}_{-0.19})\times 10^{-5}\,{\rm eV^2}\, \cr 
&&\hspace{3.3cm}{\rm and}\tan^2\theta_{12}=0.446^{+0.030}_{-0.029}\cr
{\rm for}&&\,\,\sin^2\theta_{13} < 0.04: \delta m_{23}^2=(2.1^{+0.5}_{-0.2})\times 10^{-3}\,{\rm eV^2}\cr
&& \hspace{3.1cm}{\rm and} \sin^2\theta_{23}=0.50^{+0.083}_{-0.093}
\eary

\subsection{In Vacuum}
Neutrino oscillation in vacuum would arise if neutrino were massive and mixed. For massive neutrinos, the weak eigenstates $\nu_\alpha$ are linear combinations of mass eigenstates {\small $\ket{\nu_\alpha} = \sum^n_{i=1} U^*_{\alpha i} \ket{\nu_i}$}.  After traveling a distance $L\simeq ct$, a neutrino produced with flavor $\alpha$ evolve to {\small $\ket{\nu_\alpha (t)} = \sum^n_{i=1} U^*_{\alpha i} \ket{\nu_i (t)}$}. Then, the transition probability  from a flavor estate $\alpha$ to a flavor state $\beta$ can be written as ${\small P_{\nu_\alpha\to\nu_\beta} =\delta_{\alpha\beta}-4 \sum_{j>i}\,U_{\alpha i}U_{\beta i}U_{\alpha j}U_{\beta i}\,\sin^2 [\delta m^2_{ij} L/(4\, E_\nu )]}$ with $L$ is the distance traveled by the neutrino in reaching Earth (detector).  Using the set of parameters given in eq. (\ref{3parosc}) and averaging the sine term in the probability to $\sim 0.5$ for larger distances L (longer than the solar system) \citep{lea95}, the probability matrix for a neutrino flavor vector of ($\nu_e$, $\nu_\mu$, $\nu_\tau$)$_{source}$ changing to a flavor vector  ($\nu_e$, $\nu_\mu$, $\nu_\tau$)$_{Earth}$ is given as \citep{2016JHEAp...9...25F, 2015MNRAS.450.2784F,2014MNRAS.437.2187F}
{\small
\be
{\pmatrix
{
\nu_e   \cr
\nu_\mu   \cr
\nu_\tau   \cr
}_{E}}
=
{\pmatrix
{
0.534143	  & 0.265544	  & 0.200313\cr
 0.265544	  & 0.366436	  &  0.368020\cr
 0.200313	  & 0.368020	  & 0.431667\cr
}}
{\pmatrix
{
\nu_e   \cr
\nu_\mu   \cr
\nu_\tau   \cr
}_{S}}
\label{matrixosc}
\ee
}
where $E$ and $S$, are Earth and source, respectively.

\subsection{In Matter}

\subsubsection{Two-Neutrino Mixing}

The evolution equation for the propagation of neutrinos in the above medium is given by \citep{2018PASP..130l4201F,2018PhRvD..98h3012F}
\be
i
{\pmatrix {\dot{\nu}_{e} \cr \dot{\nu}_{\mu}\cr}}
={\pmatrix
{V_{eff}-\Delta \cos 2\theta & \frac{\Delta}{2}\sin 2\theta \cr
\frac{\Delta}{2}\sin 2\theta  & 0\cr}}
{\pmatrix
{\nu_{e} \cr \nu_{\mu}\cr}},
\ee
where $\Delta=\delta m^2/2E_{\nu}$, $V_{eff}$ is the potential difference between $V_{\nu_e}$ and $V_{\nu_{\mu, \tau}}$,    $E_{\nu}$ is the neutrino energy and $\theta$ is the neutrino mixing angle.  Here we have considered the neutrino oscillation process $\nu_e\leftrightarrow \nu_{\mu, \tau}$. The transition probability in matter and as a function of $t$ 
\be
P_{\nu_e\rightarrow {\nu_{\mu}{(\nu_\tau)}}}(t) = 
\frac{\Delta^2 \sin^2 2\theta}{\omega^2}\sin^2\left (\frac{\omega t}{2}\right
),
\label{prob}
\ee
with
\be
\omega=\sqrt{(V_{eff}-\Delta \cos 2\theta)^2+\Delta^2 \sin^2
    2\theta}.
\ee
has an oscillatory behavior, with oscillation length given by
\be
L_{osc}=\frac{L_v}{\sqrt{\cos^2 2\theta (1-\frac{V_{eff}}{\Delta \cos 2\theta}
    )^2+\sin^2 2\theta}},
\label{osclength}
\ee
where $L_v=2\pi/\Delta$ is the vacuum oscillation length.  Satisfying the resonance condition
\be
V_{eff} -  \frac{\delta m^2}{2E_{\nu}} \cos 2\theta = 0,
\label{reso}
\ee
the resonance length can be written as
\be
L_{res}=\frac{L_v}{\sin 2\theta} 
\ee
\subsubsection{Three-Neutrino Mixing}
The dynamics for this case is determined by the evolution equation in a three-flavor framework which can be written as
\be
i\frac{d\vec{\nu}}{dt}=H\vec{\nu},
\ee
and the state vector in the flavor basis is defined as
\be
\vec{\nu}\equiv(\nu_e,\nu_\mu,\nu_\tau)^T.
\ee
The effective Hamiltonian is
\be
H=U\cdot H^d_0\cdot U^\dagger+diag(V_{eff},0,0),
\ee
with
\be
H^d_0=\frac{1}{2E_\nu}diag(-\delta m^2_{21},0,\delta m^2_{32}).
\ee
Here $V_{eff}$ is the effective potential calculated in section III and $U$ is the three-neutrino mixing matrix given by \citep{gon03,akh04,gon08,gon11}.  The  oscillation length of the transition probability  is given by
\be
l_{osc}=\frac{l_v}{\sqrt{\cos^2 2\theta_{13} (1-\frac{2 E_{\nu} V_e}{\delta m^2_{32} \cos 2\theta_{13}}
    )^2+\sin^2 2\theta_{13}}},
\label{osclength}
\ee
where $l_v=4\pi E_{\nu}/\delta m^2_{32}$ is the vacuum oscillation length. The resonance condition and resonance length are,
\be\label{reso3}
V_{eff}-5\times 10^{-7}\frac{\delta m^2_{32,eV}}{E_{\nu,MeV}}\,\cos2\theta_{13}=0
\ee
and
\be
l_{res}=\frac{l_v}{\sin 2\theta_{13}}.
\ee
Considering the adiabatic condition at the resonance, we  can express it as 
{\small
\be
\kappa_{res}\equiv  \frac{2}{\pi}
\left ( \frac{\delta m^2_{32}}{2 E_\nu} \sin 2\theta_{13}\right )^2
\left (\frac{dV_{eff}}{dr}\right)^{-1} \ge 1\,,
\label{adbcon}
\ee
}
where  V$_{eff}$ is given in section II. 
%

\section{Results}

We have derived the neutrino effective potential for an electron and/or positron plasma embedded in nuclear matter or SQM with a strong magnetic field.  We have considered the effective potential for the three cases shown in section II. Figures (\ref{Veff_equilibrium}), (\ref{Veff_only_electron}) and (\ref{Veff_only_positron})
show  the effective potential as a function of magnetic field, temperature and neutrino propagation angle for the transition region, SQM$^+$ (only electrons) and SQM$^-$ (only positrons), respectively.  For these plots we have considered a neutrino energy of 1 MeV, chemical potential $\mu=1$ MeV and the values of temperature, magnetic field and neutrino propagation angle listed in table~\ref{tab:1}. Columns 2, 3 and 4 in this table show the effective potential transition region, SQM$^+$ (only electrons) and SQM$^-$ (only positrons), respectively. In these figures it is shown that the neutrino effective potential is strongly dependent on the angle between the magnetic field and the neutrino propagation.

\begin{center}\renewcommand{\arraystretch}{1.5}\addtolength{\tabcolsep}{-6pt}
\begin{center}\label{tab:1}
\scriptsize{\textbf{Neutrino effective potential for an electron and positron background with ranges of  temperature,  magnetic  field  and  chemical  potential.}} \\
\end{center}
\begin{tabular}{c c c c}
 \hline \hline
\scriptsize{Quantity} &\hspace{0.3cm}  \scriptsize{$V_{\rm eff}$ (eV)}     & \hspace{0.3cm} \scriptsize{$-V_{\rm eff}$ (eV)}& \hspace{0.3cm} \scriptsize{$V_{\rm eff}$ (eV)}  \\ 
\scriptsize{} & \hspace{0.3cm}  \scriptsize{($\bar{N^0_e}\simeq N^0_e$)}  &\hspace{0.3cm}  \scriptsize{($\bar{N^0_e}\simeq 0$)} & \hspace{0.3cm}  \scriptsize{($N^0_e\simeq0$)}\\ 
\hline \hline
\scriptsize{$0.1 \leq \frac{T}{MeV} \leq 1$}	        & \hspace{0.4cm} \scriptsize{$10^{-12.5} - 10^{-10.5} $}             &\hspace{0.4cm}  \scriptsize{$10^{-7.6} - 10^{-4.8}  $} & \hspace{0.4cm} \scriptsize{$10^{-7.6} - 10^{-4.6} $}                        \\
\scriptsize{$10^3 \leq \Omega_B \leq 10^4$}     	        &\hspace{0.4cm} \scriptsize{$10^{-6.7} - 10^{-2.1}$}		&\hspace{0.4cm} \scriptsize{$10^{-6.9} - 10^{-2.0} $} &	\hspace{0.34cm} \scriptsize{$10^{-8.7} - 10^{-4.0} $}             							                     \\
\scriptsize{$0^\circ\leq \varphi \leq 90^\circ$}   	        &\hspace{0.4cm} \scriptsize{$10^{-8.1} - 10^{-2.1}$}		&\hspace{0.4cm}  \scriptsize{$10^{-8.3} - 10^{-2.3}$}	\hspace{0.4cm} & \scriptsize{$10^{-10.2} - 10^{-3.9} $}             						                     \\
\hline \hline
\end{tabular}
\end{center}
From the resonance conditions (eqs. \ref{reso} and  \ref{reso3}), we plot the contour lines of temperature and chemical potential as a function of angle for which neutrinos and anti-neutrinos can oscillate resonantly as shown in Figures \ref{ResCond_only_equilibrium}, \ref{ResCond_only_electrons} and \ref{ResCond_only_positrons}.  In these figures one can see that the temperature has two different behaviors as a function of chemical potential;  a constant function up to T $\sim$ 100 keV and an increasing function when T $\gtrsim$ 100 keV depending on the neutrino mixing parameters and the strength of the magnetic field.   For instance, when we consider Figure \ref{ResCond_only_positrons}  the temperature as a function of chemical potential starts increasing at $\mu \sim$ 251 keV, 632 keV and 858 keV for $\Omega_B=10^1$, $10^2$ and $10^3$, respectively; and accelerator parameters, and chemical potential start increasing at $\mu \sim$ 833 keV, 1051 keV and 1245 keV for $\Omega_B=10^1$, $10^2$ and $10^3$, respectively, and solar parameters. Figure \ref{ResCond_only_electrons} shows similar behaviour for the electron background regime in all curves,  the main difference being a change in the slope after $\mu \gtrsim$ 1 MeV.

Using typical Pulsar distances between $1\,{\rm kpc} \lesssim d_z \lesssim 10\,{\rm kpc}$,  average neutrino energies between  $1\,{\rm MeV} \lesssim E_{\bar\nu_e} \lesssim 10\,{\rm MeV}$, neutrino luminosities between $10^{37}\,{\rm erg\, s^{-1}} \lesssim L_{\bar\nu_e} \lesssim 10^{40}\,{\rm erg\, s^{-1}}$ and observation times between $1\,{\rm yr} \lesssim t \lesssim 10\,{\rm yr}$, the number of expected events detected by the SK (dotted green line), HK (dotted-dashed blue line) and DUNE (dashed orange line) experiments coming from the SQM$^+$--SQM$^-$ region inside a strange star are plotted in Figure \ref{events}.    The upper left-hand panel corresponds to the number of events as a function of distance for $L_{\bar{\nu}_e}=10^{39}\:\mathrm{erg\, s^{-1}}$, $ t=1\, {\rm yr}$ and $E_{\bar\nu_e} = 1\,{\rm MeV}$. This panel displays that the number of  events lie in the range of $4\times 10^{-3}$ and $1.1\times 10^2$. The upper right-hand panel shows the number of events as a function of the average neutrino energy  for $L_{\bar{\nu}_e}=10^{39}\:\mathrm{erg\, s^{-1}}$, $t=1\, {\rm yr}$ and $d_z = 1\,{\rm kpc}$. This panel exhibits that the neutrino events range from $0.15$ to $48.2$. The lower left-hand panel exhibits the number of events as a function of the neutrino luminosity for $d_z =1\,{\rm kpc}$, $ t=1\, {\rm yr}$ and $E_{\bar\nu_e} = 1\,{\rm MeV}$. This panel exhibits that the neutrino events per year range from $6\times 10^{-3}$ to $56.6$.  The lower right-hand panel displays the number of events as a function of observation time for $L_{\bar{\nu}_e}=10^{39}\:\mathrm{erg\, s^{-1}}$, $d_z = 1\,{\rm kpc}$ and $E_{\bar\nu_e} = 1\,{\rm MeV}$. This panel shows that the number of  events per year lies in the range of $5\times 10^{-2}$ and $54.3$.\\

It can be observed that the neutrino events coming from the electric-charge phase transition could be detected in the new generation of neutrino experiments such as HK, especially in those pulsar located at distances of a few kpc. It is worth noting that our estimates were done for one source. If we consider the large pulsar population, the number of neutrinos would  increase dramatically.\\

Figure \ref{probability} shows the oscillation probabilities of neutrino events as a function of neutrino energy for $B=10^2\,B_c$, $L=1\, {\rm km}$ and $\varphi=0^\circ$ (upper left-hand panel), $\varphi=30^\circ$ (upper right-hand panel), $\varphi=60^\circ$ (lower left-hand panel) and $\varphi=90^\circ$ (lower right-hand panel). This figure displays that electron neutrino can hardly oscillate to muon and tau neutrinos, and muon and tau neutrinos oscillate resonantly between them.  These panels show that the oscillation probabilities depend strongly on the angles between the neutrino propagation and the magnetic field.  This implies that neutrinos in such a strong magnetic field could provide information on its topology. 

\section{Discussion and Conclusions}
First, we would like to point out that there exist several papers in the literature where calculations involving magnetic fields as large as $10^{20}$ G are employed in neutron or quark stars.  One has to be careful with this, as such a large magnetic field cannot be produced or maintained by a NS; and if it did, the mass of such an object would be on the order of a few thousand solar masses, i.e., it would either, destroy the star before achieving such magnitude, or collapse into a BH before that density was reached.
Following the ideas in \citet{1953ApJ...118..116C},  \citet{1991ApJ...383..745L} estimated that the magnetic field may not exceed some $10^{18}$ G, which already amounts for around $10\%$ of the mass of a typical NS (i.e., the binding energy).  We have, therefore, restricted the magnetic fields in this study to this limit. 

We have calculated, for the first time, the neutrino effective potential for an electron-only background and positron-only background when a large magnetic field is present in a neutron or quark star.  We find that the neutrinos oscillate resonantly considering the best-fit values of the two-neutrino mixing and three-neutrino mixing.

The electric-phase-transition region studied in this paper allows for a different cooling mechanism in strange stars.  
A few tens of seconds after CC, the compact star cools below 1 MeV and the energy (per event) released by neutrinos from e$^-$ -- e$^+$ pair annihilation could become a signature against the thermal-neutrinos (and anti-neutrinos) background from the URCA process.
Now, the e$^-$s and e$^+$s must be constantly created (by these URCA processes) on both sides of the electric interface to provide the supply of pairs that annihilate. 
Otherwise, if the supply cannot be met by the URCA processes, a charge starts building up (on each side of the SQM$^-$ -- SQM$^+$ interface) such that the SQM$^-$ region repels e$^-$s and the SQM$^+$ repels e$^+$s and the mechanism shuts down.


Losing energy at a rate of $\sim 1$ MeV per (anti) neutrino, besides the thermal URCA neutrinos should provide for cooling timescales which are considerably shorter than the usual channels.

Computing the neutrino effective potential for the most general case, we find that the Fermion distribution functions for electrons and positrons, which depend on temperature, chemical potential, and magnetic field, are not soluble for the condition $E_e\lesssim \mu_e$ (see Appendix~ A.).

An important result from this work is the neutrino oscillation dependence on the angle between the neutrino propagation and the magnetic field.  
Given that all charged particles, are confined to Landau levels, in particular e$^-$s and e$^+$s (in fact, what produces the $V_{eff}$ over neutrinos is the interaction of the magnetic field with W$^\pm$ gauge bosons), then their momenta will be confined to the direction (both parallel and/or anti-parallel) of the magnetic-field lines, thus, in an annihilation event, the resulting neutrino -- antineutrino pair has to conserve momentum and thus the neutrinos themselves will travel, preferentially, along the direction of the (local) magnetic field lines.
This implies that neutrinos in such a strong magnetic field could provide information of the topology of the field inside the star.
Furthermore, in a rotating compact star this implies that the neutrino flux will depend on the phase of the spin.  
Thus, in principle, one could detect neutrino pulses of rotating magnetars with electric phase transition around 1 MeV.  
Nonetheless, this will depend on the magnetar being close enough to the detector such that the flux is large enough.
A good exercise is to look for times of arrival of detected neutrinos in Super-Kamiokande and look for $\sim 1$ MeV neutrinos which may have arrived at multiples of the same period.  
Were this cooling mechanism to persist beyond the point where the magnetar is beyond the "death line" (no longer producing EM pulses), their periods could be longer than several seconds.
Also, the internal configuration of the field may strongly differ from the external (dipolar) one, thus the neutrino flux pulsation may not be in phase with the electromagnetic counterpart.

We have estimated the number of neutrino events and the standard flavor ratio from this electric-charge phase transition present in strange stars. The neutrino events expected from the hyper-accretion phase on Super-Kamiokande, Hyper-Kamiokande and DUNE experiments range from $10^{-3}$ to 100 events. which exhibit a nonsignificant deviation of the standard flavor ratio (1:1:1).


The gravitational-wave transient GW170817 was detected at 2:41:04 UTC, 2017 August 17 by LIGO and Virgo experiments \citep{PhysRevLett.119.161101,2041-8205-848-2-L12}.  This event, associated with a binary neutron star merger,  triggered immediately the Gamma-ray Burst Monitor (GBM) onboard Fermi Gamma-ray Space Telescope at 12:41:06 UTC \citep{2017ApJ...848L..14G} and after the INTErnational Gamma-Ray Astrophysics Laboratory (INTEGRAL) \citep{2017ApJ...848L..15S}. The electromagnetic counterpart, called  GRB 170817A, followed up  by multiple ground-based telescopes in several wavelength bands \citep{troja2017a, 2017ApJ...848L..21A,  2017ATel11037....1M}  has been associated to different emission mechanisms in GRB afterglows \citep{2017arXiv171005896G, 2017arXiv171005897B, 2019ApJ...871..200F, 2019ApJ...871..123F, 2018arXiv180207328V}. Depending on the nuclear EoS, it may be that one or two of the binary components are not NSs but SSs.  If this were the case, the merger would heat up the SQM inside the star(s), increasing the neutrino flux
and activating the neutrino-flavor oscillations we have described throughout this study.
%


\section*{Acknowledgements}
\label{sec-Ack}

We thank Dany Page for useful discussions.  NF acknowledges financial support from UNAM-DGAPA-PAPIIT through grant IA102917.  This research has made use of NASA’s Astrophysics Data System as well as arXiv.


\clearpage

\section*{Appendix A: }

{\bf Fermi Dirac} (FD) integrals
\begin{equation}
I_n = \int_0^\infty \frac{z^{n-1}}{e^z+1}dz, \qquad n>1
\end{equation}
It is easy to prove that
\begin{equation}\label{eqn:FD}
I_n = \left(1-2^{1-n} \right)\Gamma\left(n\right)\zeta\left(n\right)
\end{equation}
Where $\zeta(n)$ is Riemann's {\it zeta} function and $\Gamma(n)$ is the {\it gamma} function.

Now, to solve {\bf Sommerfeld} (S) integrals:
\begin{equation}
	I_S=\int_0^\infty \frac{f(\epsilon)}{e^{\beta(\epsilon-\mu)}+1}d\epsilon
\end{equation}
Allowing $z=\beta(\epsilon-\mu)$, then $d\epsilon=dz/\beta$, and so
\begin{eqnarray*}
	I_S &=& \frac{1}{\beta}\int_{-\beta\mu}^\infty \frac{f(z/\beta + \mu)}{e^z + 1}dz \\
	  &=& \frac{1}{\beta}\int_{-\beta\mu}^0\frac{f(z/\beta + \mu)}{e^z + 1}dz +
	  	  \frac{1}{\beta}\int_{0}^\infty \frac{f(z/\beta + \mu)}{e^z + 1}dz \\
	  &=& \frac{1}{\beta}\int_0^{\beta\mu} f(-z/\beta + \mu)\left(1-\frac{1}{e^{z} + 1}\right)dz \\
	  &&\hspace{4cm} +\frac{1}{\beta}\int_{0}^\infty \frac{f(z/\beta + \mu)}{e^z + 1}dz \\
	  &=& \frac{1}{\beta}\int_0^{\beta\mu} f(-z/\beta + \mu)dz\\     &&\hspace{2cm} -\frac{1}{\beta}\int_0^{\beta\mu}\frac{f(-z/\beta + \mu)}{e^{z} + 1}dz\\ 
	  &&\hspace{4cm} +\frac{1}{\beta}\int_{0}^\infty \frac{f(z/\beta + \mu)}{e^z + 1}dz 
\end{eqnarray*}

On the first term we rename the variable $x=-\frac{z}{\beta}+\mu$, then $dz= -\beta dx$, and the integral runs from $\mu$ to 0. So, the first term is exactly $\int_0^\mu f(x)dx$.

For the second term, if we allow $\beta\mu \gg 1$, then we can consider the second term as if it was running from 0 to $\infty$. Then, $I_S$ can be simplified even more:
\begin{equation}
I_S = \int_0^\mu f(x)dx + \frac{1}{\beta} \int_0^\infty\frac{f(z/\beta+\mu)-f(-z/\beta+\mu)}{e^z+1}dz
\end{equation}

Expanding the function $f$ as a Taylor series, that is\\ $f(x)=\sum_{n=0}^\infty \frac{f^{(n)}(a)}{n!}\left(x-a\right)^n$, we obtain
\begin{eqnarray}
  f(z/\beta+\mu) &=& \sum_{n=0}^\infty \frac{f^{(n)}(\mu)}{n!}\left(\frac{z}{\beta}\right)^n \\
  f(-z/\beta+\mu) &=& \sum_{n=0}^\infty \frac{f^{(n)}(\mu)}{n!}(-1)^n\left(\frac{z}{\beta}\right)^n 
\end{eqnarray}

This way
\begin{eqnarray*}
I_S &=& \int_0^\mu f(x)dx\\
&&\hspace{0.5cm}+ \frac{1}{\beta} \int_0^\infty\frac{2\sum_{n=0}^\infty \frac{f^{(2n+1)}(\mu)}{(2n+1)!}\left(\frac{z}{\beta}\right)^{2n+1}}{e^z+1}dz \\
 &=& \int_0^\mu f(x)dx + 2\sum_{n=0}^\infty \frac{f^{(2n+1)}(\mu)}{(2n+1)!\left(\beta\right)^{2n+2}}\int_0^\infty\frac{z^{2n+1}}{e^z+1}dz \\
 &=& \int_0^\mu f(x)dx + 2\sum_{n=0}^\infty \frac{f^{(2n+1)}(\mu)}{(2n+1)! \left(\beta\right)^{2n+2}} \\
 &&\hspace{2cm}\times\left(1-2^{-(2n+1)}\right)\Gamma(2n+2) \zeta(2n+2)
\end{eqnarray*}



In this work we need to solve the following integral 
\begin{equation}\label{eqn:Integral}
	I = \int_{0}^{\infty}E_{e,n}^m (f_{e,n}+\overline{f}_{e,n}) dP_3 
\end{equation}
for $m=-1,0,1$, and $$f_{e,n}=\frac{1}{e^{\beta(E_{e,n}-\mu)}+1},\qquad \overline{f}_{e,n}=\frac{1}{e^{-\beta(E_{e,n}+\mu)}+1}$$where$$E_{e,n}^2=P_3^2+m_e^2+H.$$

For the sake of simplicity, $w=E_{e,n}$, and let us integrate with respect to $w$:

\begin{equation}
	I = \int_{A}^\infty \left(\frac{1}{e^{\beta(w-\mu)}+1}+\frac{1}{e^{-\beta(w+\mu)}+1}\right)\frac{w^{m+1}}{\sqrt{w^2-A^2}}dw
\end{equation} where $A^2=m_e^2+H$.

Now we consider $\epsilon = w-A$, then $d\epsilon=dw$, and so 
\be
I=&\int_0^\infty& \left( \frac{1}{e^{\beta(\epsilon+A-\mu)}+1}+\frac{1}{e^{-\beta(\epsilon+A+\mu)}+1}\right)\times\nonumber\\
&&\frac{(\epsilon+A)^{m+1}}{\sqrt{\epsilon^2+2\epsilon A}}d\epsilon.
\ee
Breaking this into two integrals and making $\mu_1 = \mu - A$ and $\mu_2 = \mu+A$. 
\begin{eqnarray}
I_1 &=& \int_0^\infty \frac{f(\epsilon)}{e^{\beta(\epsilon - \mu_1)}+1} d\epsilon\\
I_2 &=& \int_0^\infty \frac{f(\epsilon)}{e^{-\beta(\epsilon + \mu_2)}+1} d\epsilon
\end{eqnarray}
where $$f(\epsilon)=\frac{(\epsilon+A)^{m+1}}{\sqrt{\epsilon^2+2\epsilon A}}.$$

Both integrals can be fully solved by a combination of techniques developed further down the text. In particular, $I_1$ is a simple Sommerfeld integral, while $I_2$ can be obtained in a similar fashion. The only problem is that $I_2$ does not converge in $[0,\infty)$ for those particular values of $m$. $I_2$ converges if and only if $m<-1$.

To observe the divergence of $I_2$, we must notice that \be\frac{1}{e^{-z}+1}=1-\frac{1}{e^z+1},\ee 
which makes 
\begin{equation}
	I_2 = \int_0^\infty f(\epsilon)d\epsilon - \int_0^\infty \frac{f(\epsilon)}{e^{\beta(\epsilon+\mu_2)}+1}d\epsilon.
\end{equation}
The second term is just another application of the Sommerfeld integral. However, the first term 
\begin{equation}I_3=
\int_0^\infty f(\epsilon)d\epsilon= \int_0^\infty \frac{(\epsilon+A)^{m+1}}{\sqrt{\epsilon^2+2\epsilon A}}d\epsilon
\end{equation}
is a radical function which only converges when the overall degree of the function is less than -1. Which also can be translated as $m<-1$. 

Considering everything, $I_2$ will not converge for $m\geq -1$, and neither will $I$.

This last integral $I_3$ can actually be calculated for $m<-1$ as\begin{equation}
I_3=\int_0^\infty (y^2+A^2)^{\frac{m}{2}}dx
\end{equation}
with the substitution $y=\sqrt{\epsilon^2+2\epsilon A}$. $I_3$ can now be solved as
\begin{equation}
I_3 = \frac{A^{m+1}\sqrt{\pi}\,\Gamma\left[-\left(\frac{m+1}{2}\right)\right]}{2\Gamma\left(-\frac{m}{2}\right)}
\end{equation}
for $m<-1$, where $\Gamma$ is the {\it gamma} function.

\begin{figure*}[htp]
\centering
 \includegraphics[width=0.9\textwidth]{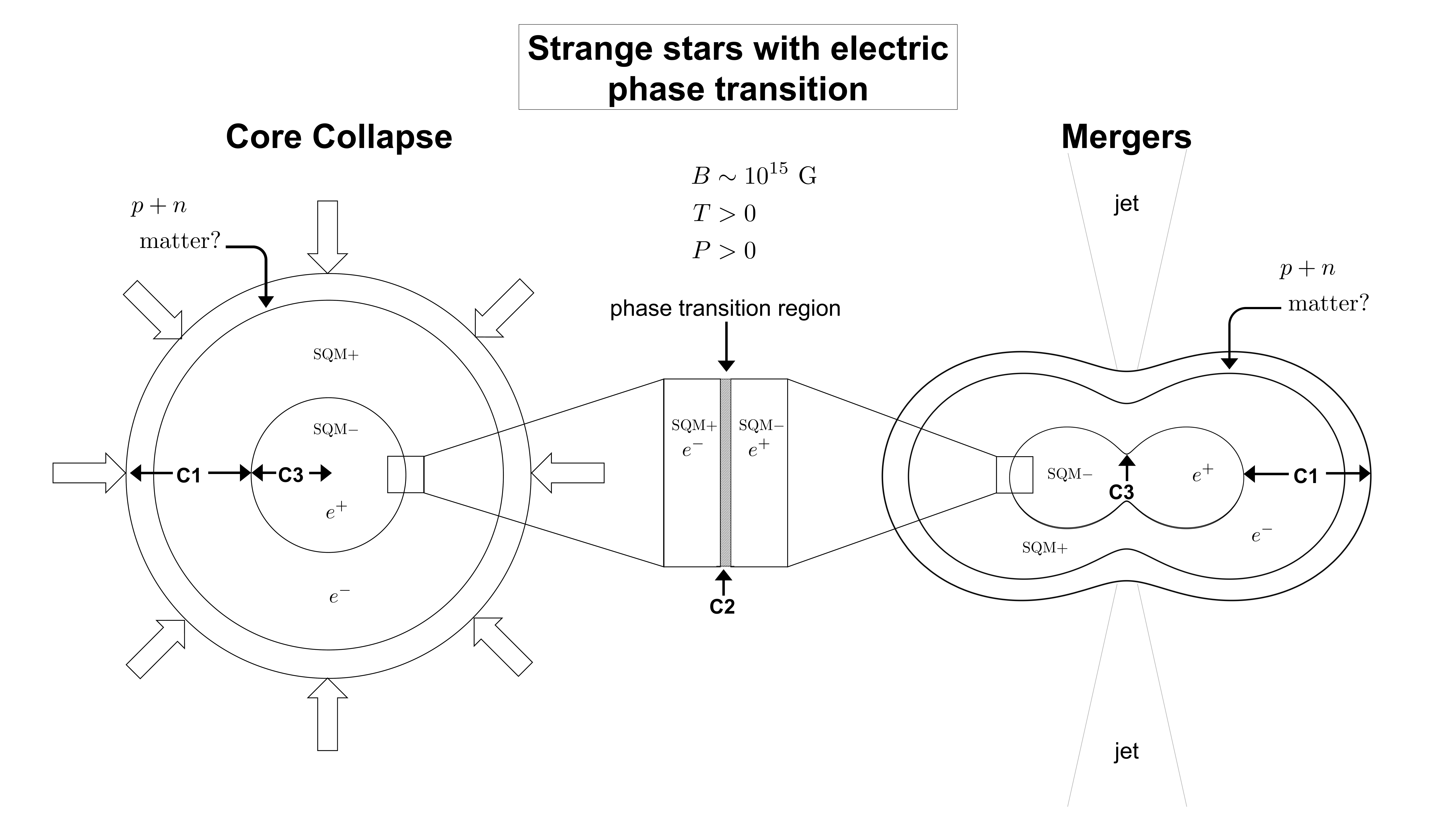}
\caption{\label{diagram}On the left-hand side of the diagram, we show an schematic view of a strange star with the electric phase transition between SQM$^+$ and SQM$^-$ during core collapse.  The region around the phase transition (C2) is zoomed at the central part of the diagram.  The right-hand side of the diagram illustrates the same C2 region during the merger of two strange stars. Region C1 may exist for SQM at zero pressure, whereas region C3, that with SQM$^-$ may only occur at finite pressure within a strange star. In the case when no electric phase transition exists, C1 would occupy the whole compact object.}
\end{figure*}

\begin{figure*}[htp]
\centering
 \includegraphics[width=0.9\textwidth]{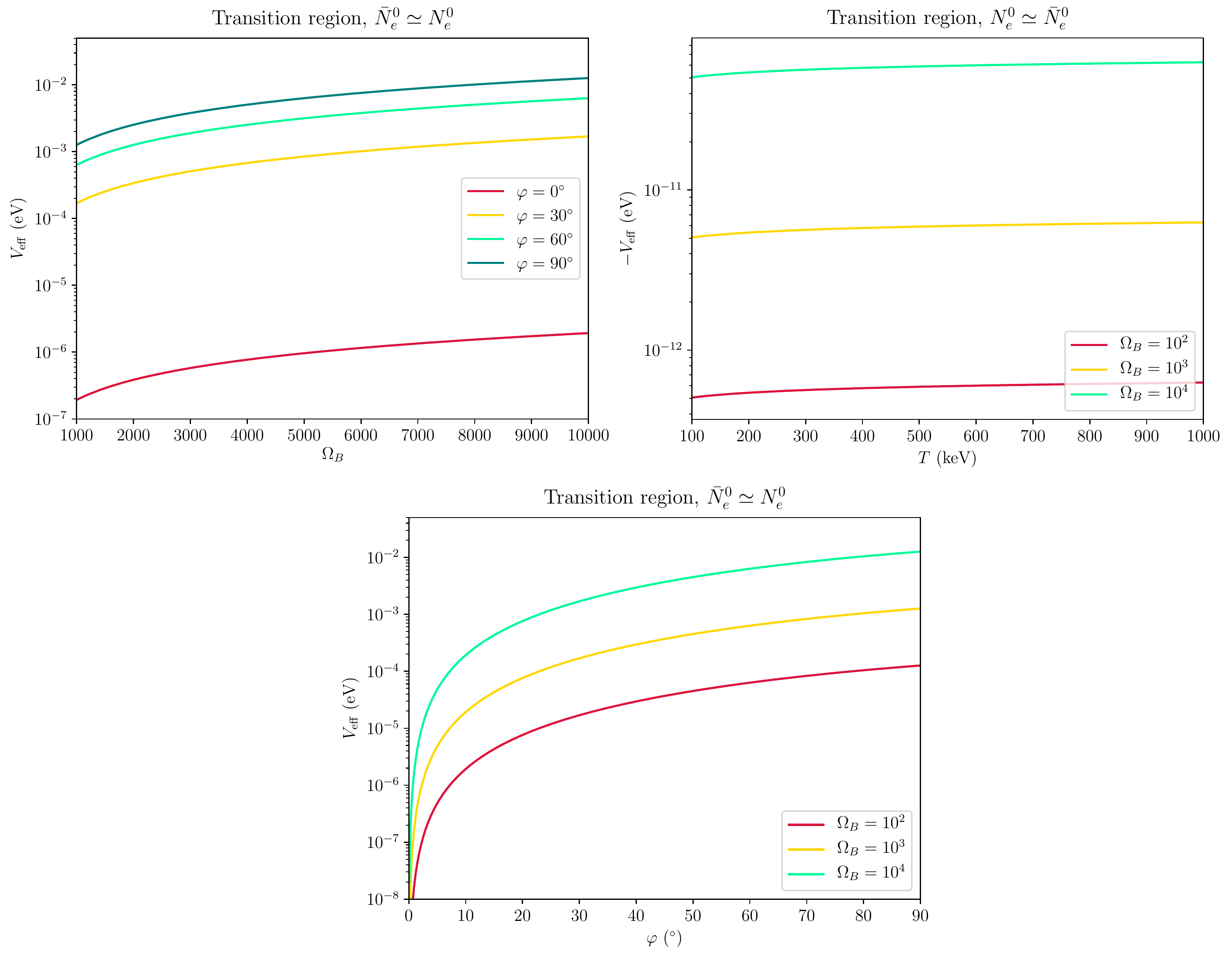} 
\caption{\label{Veff_equilibrium}  The effective potential (V$_{eff}$) as a function of magnetic field (B), temperature (T) and
the angle of propagation with respect to the magnetic field ($\varphi$) for $\bar{N}^0_e=N^0_e$}
\end{figure*}

\begin{figure*}[htp]
\centering
 \includegraphics[width=0.9\textwidth]{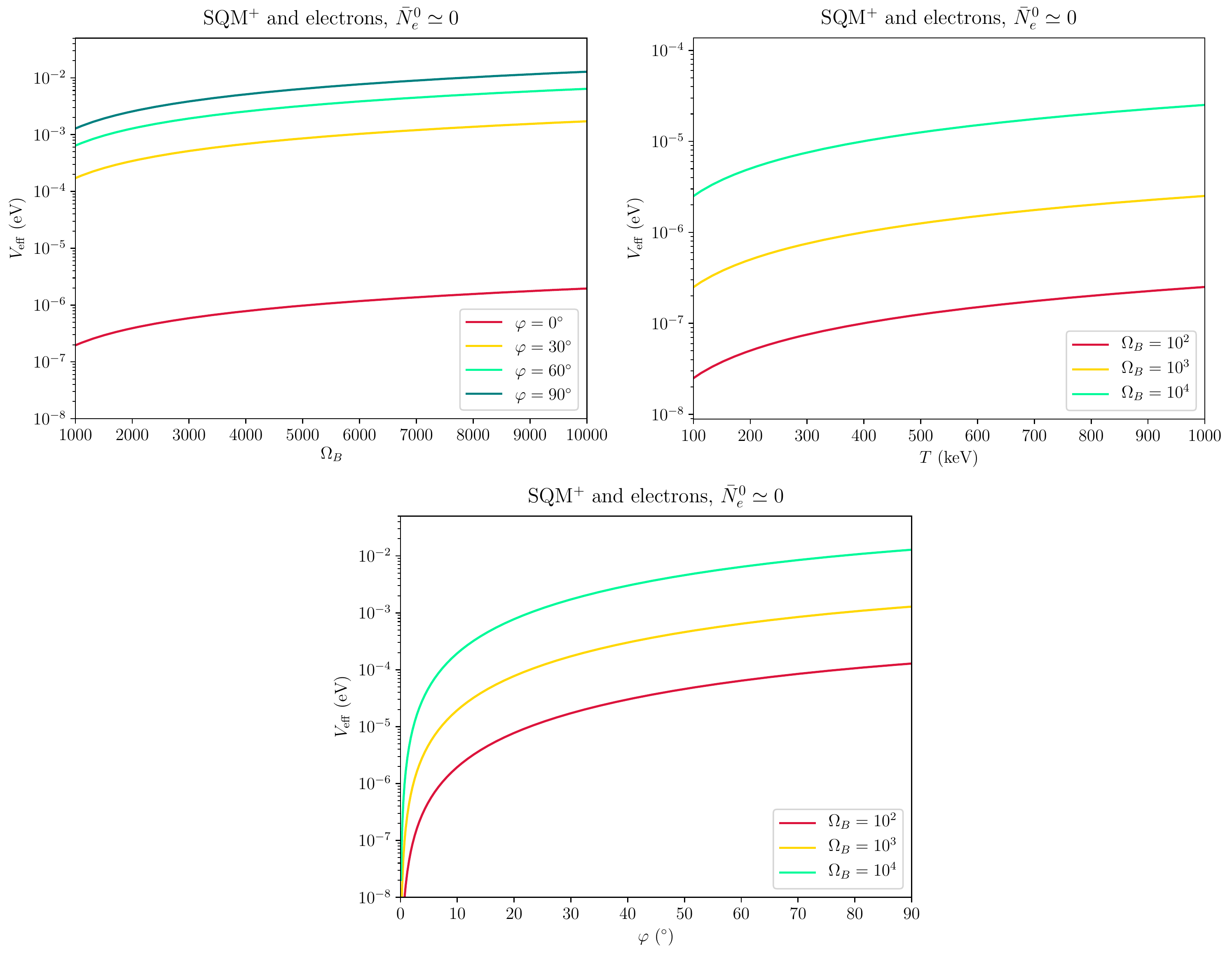} 
\caption{\label{Veff_only_electron}  The effective potential (V$_{eff}$) as a function of magnetic field (B), temperature (T) and
the angle of propagation with respect to the magnetic field ($\varphi$) for $\bar{N}^0_e=0$.}
\end{figure*}

\begin{figure*}[htp]
\centering
 \includegraphics[width=0.9\textwidth]{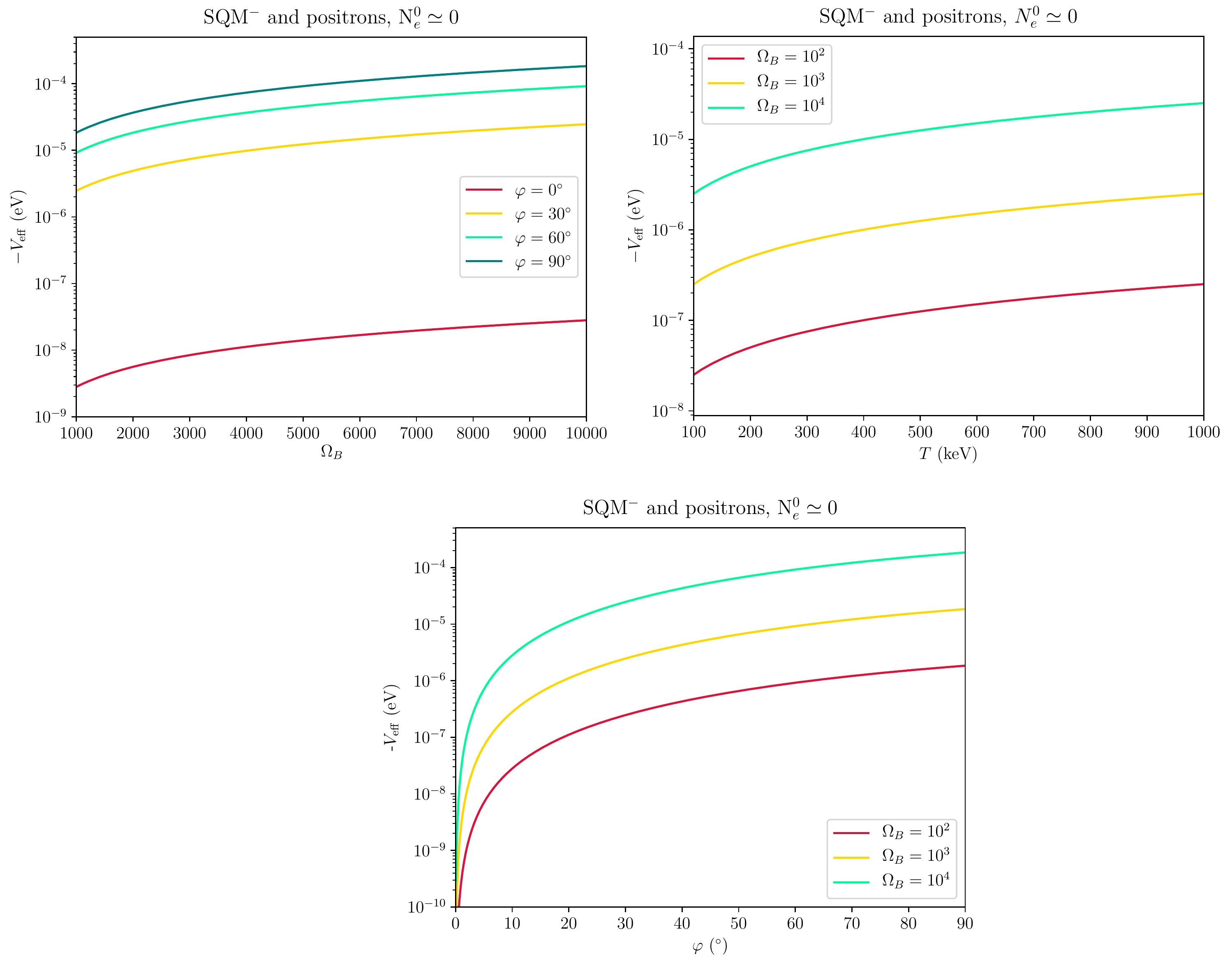} 
\caption{\label{Veff_only_positron}  The effective potential (V$_{eff}$) as a function of magnetic field (B), temperature (T) and
the angle of propagation with respect to the magnetic field ($\varphi$) for $N^0_e$=0.}
\end{figure*}


\begin{figure*}[htp]
\centering
 \includegraphics[width=\textwidth]{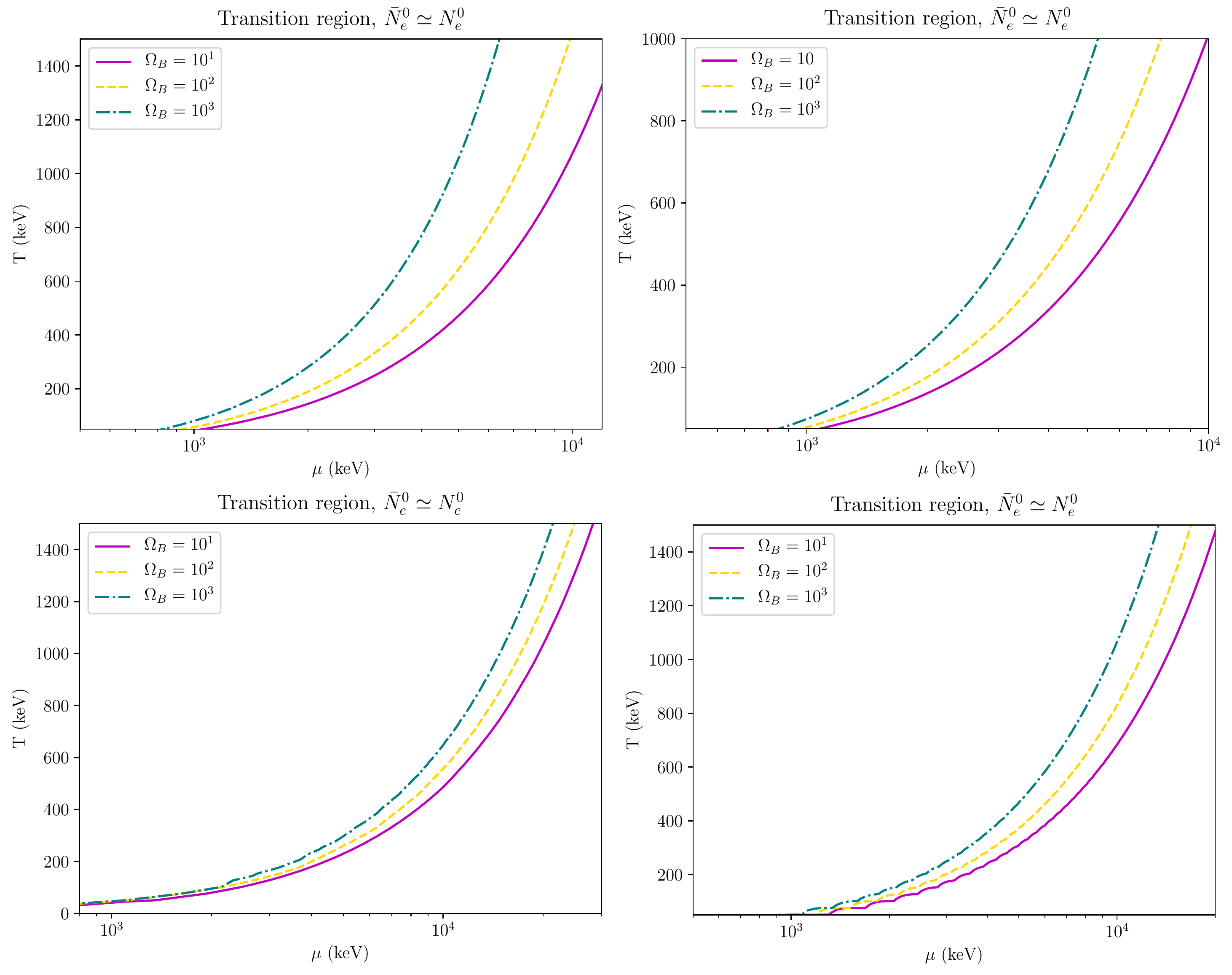} 
\caption{Contour plots of temperature  (T) and chemical potential  ($\mu$) as a function of the magnetic field  for which the resonance condition is satisfied. We have applied the neutrino effective potential  for $\bar{N}^0_e=N^0_e$,
$\mu=1 MeV$, T=500 keV, $\varphi=0^\circ$,
$E_\nu=1$ MeV  and used the best-fit values  of the two-neutrino mixing; solar (top, left-hand side figure),  atmospheric (top, right-hand side figure) and accelerator (bottom, left-hand side figure), and three-neutrino mixing (bottom, right-hand side figure).}\label{ResCond_only_equilibrium}
\end{figure*}

\begin{figure*}[htp]
\centering
 \includegraphics[width=\textwidth]{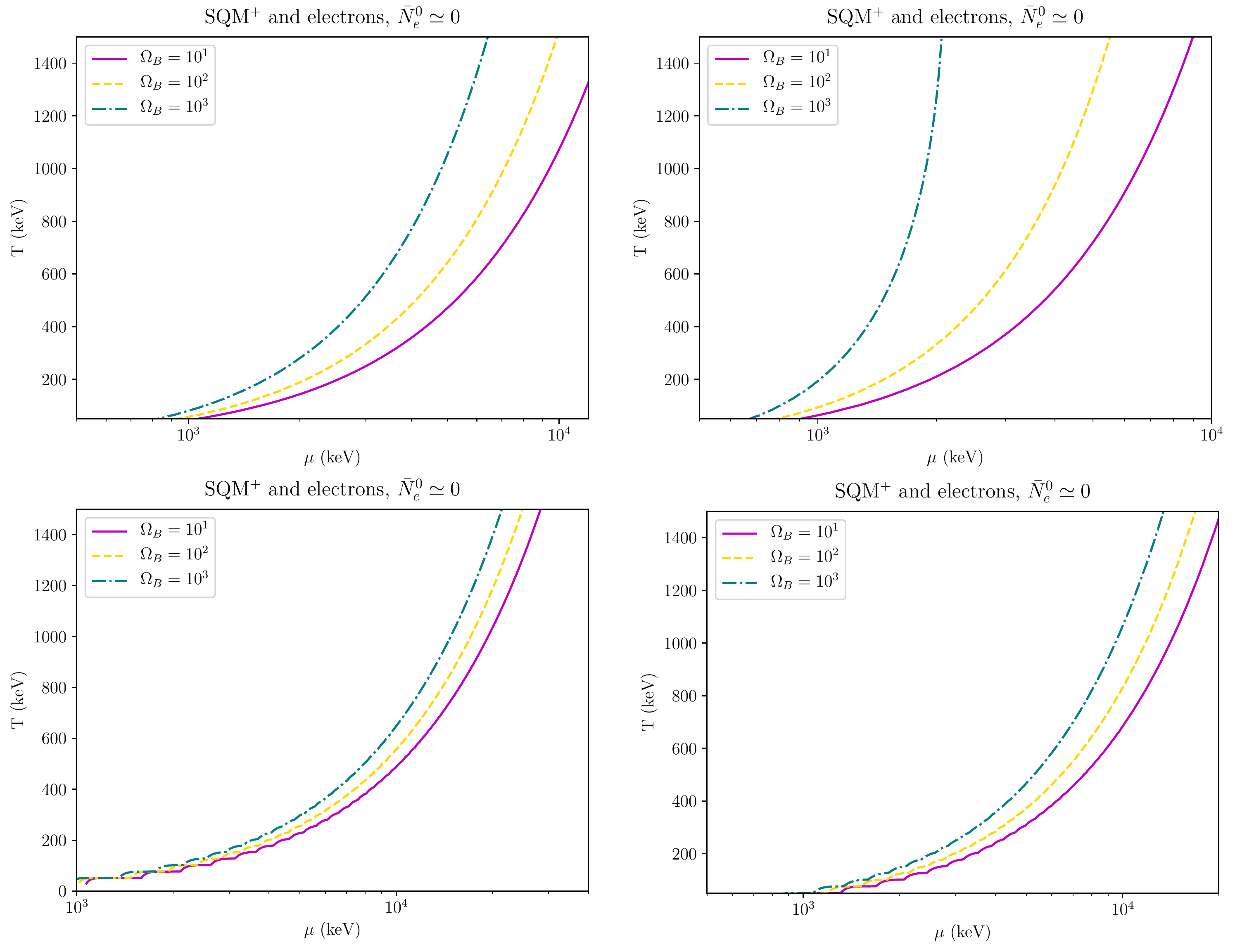} 
\caption{Contour plots of temperature  (T) and chemical potential  ($\mu$) as a function of the magnetic field  for which the resonance condition is satisfied. We have applied the neutrino effective potential  for $\bar{N}^0_e=0$,
$\mu=1$ MeV, T=500 keV, $\varphi=0^\circ$,
$E_\nu=1 MeV$  and used the best-fit values  of the two-neutrino mixing; solar (top, left-hand side figure),   atmospheric (top, right-hand side figure) and accelerator (bottom, left-hand side figure), and three-neutrino mixing (bottom, right-hand side figure).}\label{ResCond_only_electrons}
\end{figure*}

\begin{figure*}[htp]
\centering
 \includegraphics[width=\textwidth]{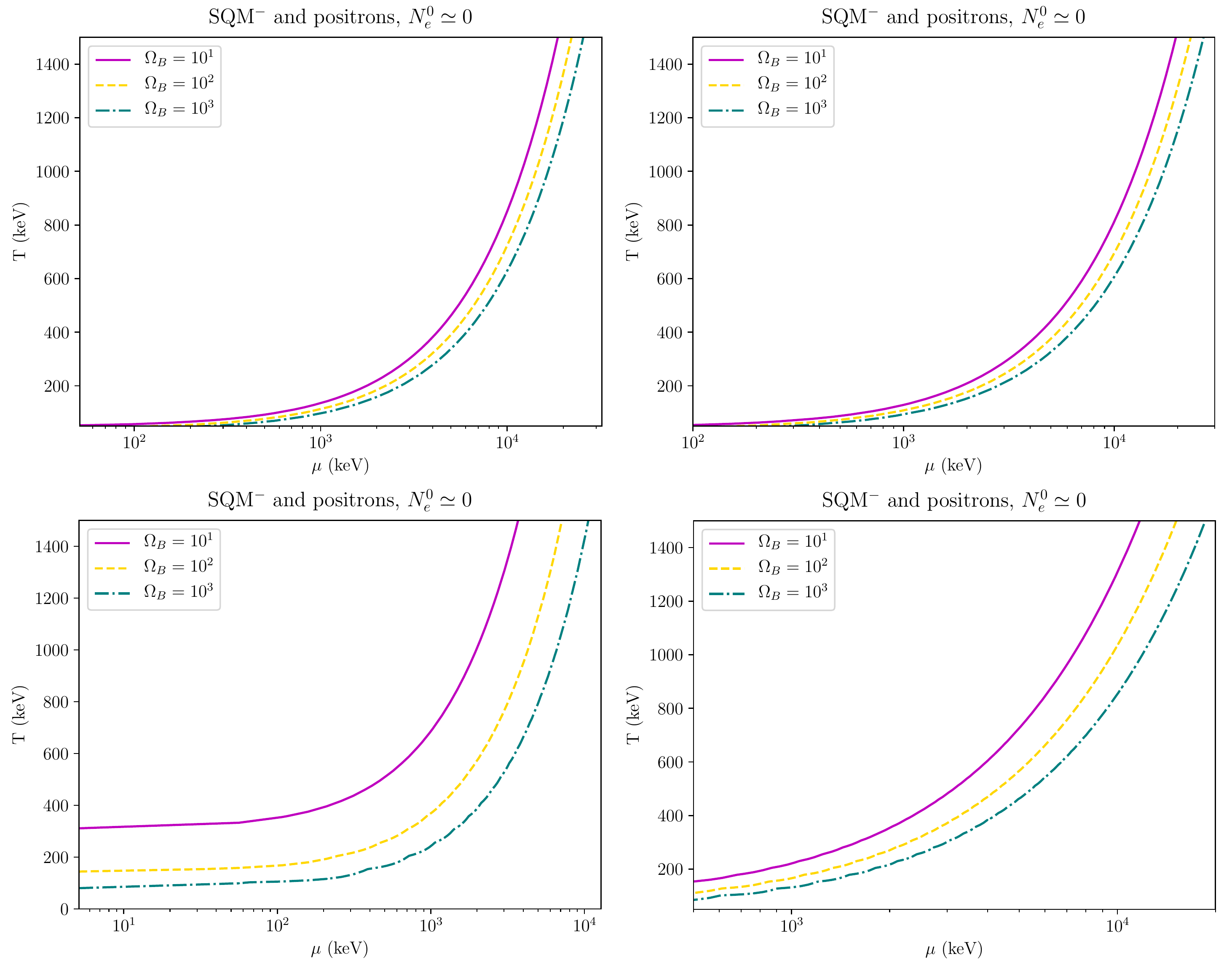} 
\caption{Contour plots of temperature  (T) and chemical potential  ($\mu$) as a function of the magnetic field  for which the resonance condition is satisfied. We have applied the neutrino effective potential  for $N^0_e=0$,
$\mu=1$ MeV, T=500 keV, $\varphi=0^\circ$,
$E_\nu=1 MeV$  and used the best-fit values  of the two-neutrino mixing; solar (top, left-hand side figure),  atmospheric (top, right-hand side figure) and accelerator (bottom, left-hand side figure), and three-neutrino mixing (bottom, right-hand side figure).}\label{ResCond_only_positrons}
\end{figure*}

\begin{figure*}[htp]
\centering
 \includegraphics[width=\textwidth]{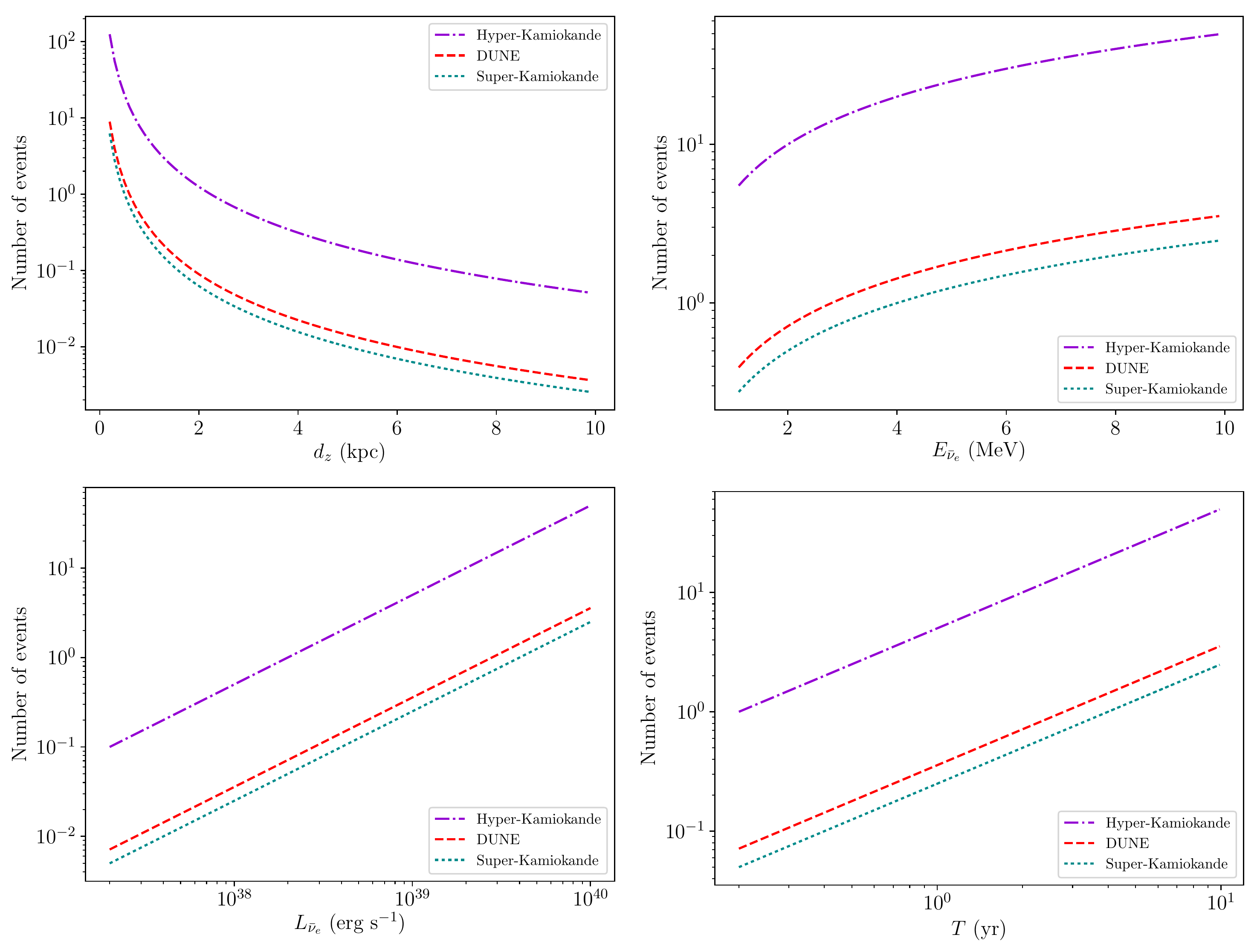} 
\caption{Number of events expected on SK (dotted green line), HK (dotted-dashed purple line) and DUNE (dashed red line).   The upper left-hand panel corresponds to the number of events as a function of distance for $L_{\bar{\nu}_e}=10^{39}\:\mathrm{erg\, s^{-1}}$, $ t=1\, {\rm yr}$ and $E_{\bar\nu_e} = 1\,{\rm MeV}$.   The upper right-hand panel shows the number of events as a function of the average neutrino energy  for $L_{\bar{\nu}_e}=10^{39}\:\mathrm{erg\, s^{-1}}$, $t=1\, {\rm yr}$ and $d_z = 1\,{\rm kpc}$.  The lower left-hand panel exhibits the number of events as a function of the neutrino luminosity for $d_z =1\,{\rm kpc}$, $ t=1\, {\rm yr}$ and $E_{\bar\nu_e} = 1\,{\rm MeV}$.   The lower right-hand panel displays the number of events as a function of observation time for $L_{\bar{\nu}_e}=10^{39}\:\mathrm{erg\, s^{-1}}$, $d_z = 1\,{\rm kpc}$ and $E_{\bar\nu_e} = 1\,{\rm MeV}$.}\label{events}
\end{figure*}

\begin{figure*}[htp]
\centering
 \includegraphics[width=\textwidth]{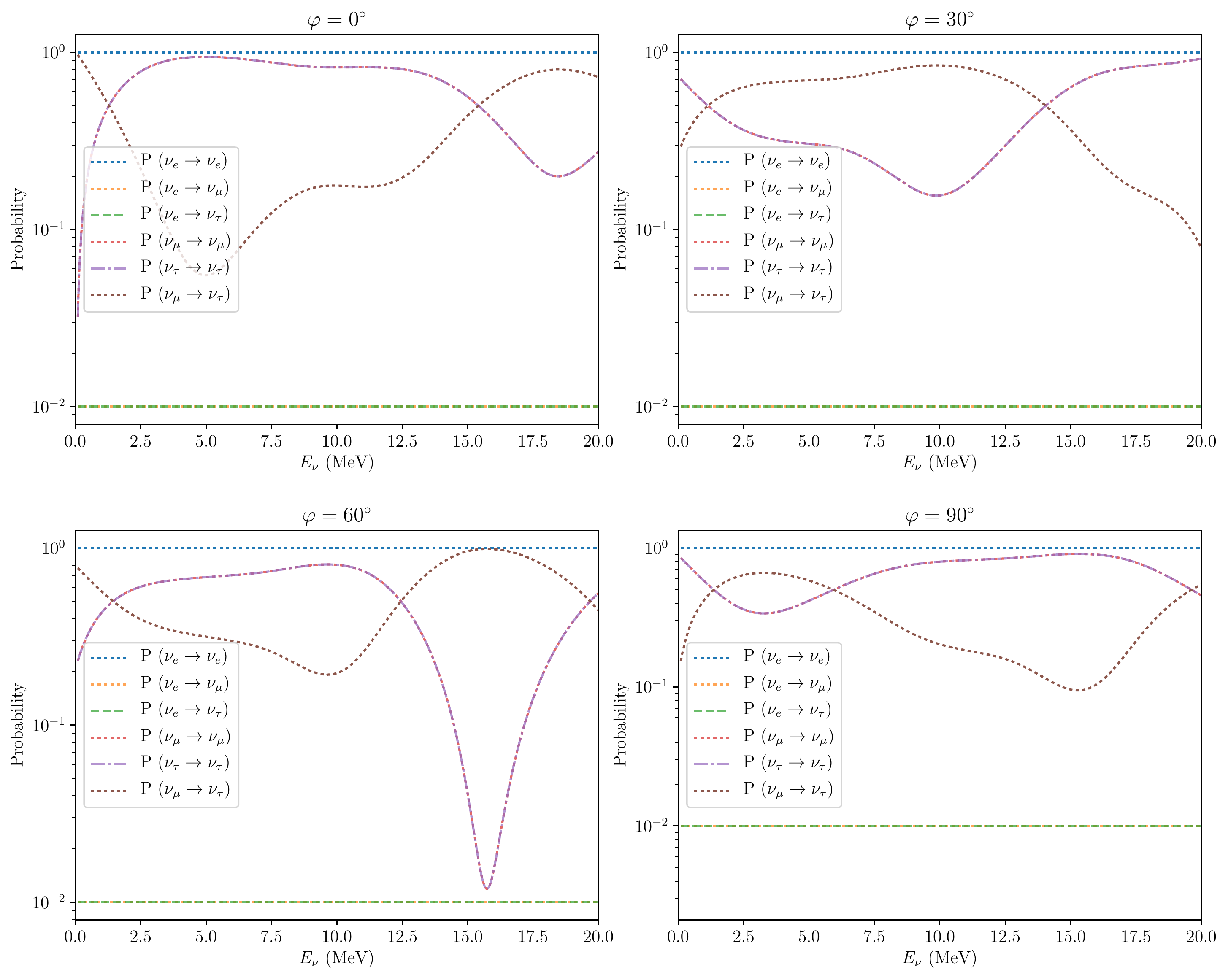} 
\caption{Oscillation probabilities of neutrino events as a function of neutrino energy for $B=10^2\,B_c$, $L=1\, {\rm km}$ and $\varphi=0^\circ$ (upper left-hand panel), $\varphi=30^\circ$ (upper right-hand panel), $\varphi=60^\circ$ (lower left-hand panel) and $\varphi=90^\circ$ (lower right-hand panel).}\label{probability}
\end{figure*}


\end{document}